\documentclass[prl,amsmath,amssymb,twocolumn,superscriptaddress]{revtex4}
\usepackage{amsmath,amssymb,graphicx}
\usepackage{dcolumn}

\begin{document}

\title{Observation of Ballistic Thermal Transport in a Nonintegrable Classical Many-Body System}

\author{Jianjin Wang}
\email{phywjj@foxmail.com}

\affiliation{Department of Physics, Jiangxi Science and Technology Normal University, Nanchang 330013, Jiangxi, China}
\affiliation{MinJiang Collaborative Center for Theoretical Physics, College of Physics and Electronic Information Engineering, Minjiang University, Fuzhou 350108, China}

\author{Daxing Xiong}
\email{xmuxdx@163.com}
\affiliation{MinJiang Collaborative Center for Theoretical Physics, College of Physics and Electronic Information Engineering, Minjiang University, Fuzhou 350108, China}

\begin{abstract}
We report, for the first time, the observation of ballistic thermal transport in a nonintegrable classical many-body system. This claim is substantiated by appropriately incorporating long-range interactions into the system, which exhibits all characteristic hallmarks of ballistic heat transport, including the presence of equilibrium dynamical correlations exhibiting ballistic scaling, a size-independent energy current and a flat bulk temperature profile. These findings hold true for large system sizes (long times), indicating that ballistic heat transport is valid in the thermodynamic limit. The underlying mechanism is attributed to the presence of traveling discrete breathers in the relevant nonintegrabel systems surpassing conventional solitons in a nonlinear integrable Toda system.
\end{abstract}

\maketitle
\emph{Introduction}.---Integrability is a fundamental concept in the analysis of classical mechanics~\cite{Book-1}. A system is considered integrable if it possesses an equal number of independent conserved quantities as degrees of freedom. Consequently, integrable systems, though being sparse in nature, represent one of the key topics in mathematical physics as they gave birth to the celebrated soliton theory~\cite{Book-2}, which provides explanations for various observable phenomena, ranging from localized light in nonlinear optics, waves on shallow water, and tsunami waves, to elementary particles and localized excitations in condensed matter at low temperatures.

From the perspective of classical statistical mechanics, integrable systems are particularly intriguing~\cite{Int-1}. They give a testing ground for the conjecture~\cite{Int-2} that the thermodynamics of a large class of systems can be precisely described in terms of quasiparticles such as solitons. These quasiparticles interact in a specific manner without exchanging energy and momentum, thereby reducing the available phase space~\cite{Int-3}, ultimately leading to ballistic thermal transport~\cite{Int-4, Int-5}. The characteristic features include ballistic scaling of equilibrium dynamical correlations, a size-independent energy current, and a flat bulk temperature profile, these results being valid in the thermodynamic limit~\cite{TypeA-1,TypeA-2,TypeA-3}.

In nonintegrable systems, where the conservation laws are absent, and interactions between quasiparticles are more complicated, such ballistic behavior is usually not observed. For a nonintegrable system, intensive studies suggested that, whether or not the system has a conserved total momentum is a key ingredient~\cite{NonInt-1,NonInt-2,NonInt-3}. This factor, combined with integrability, leads to three broad categories of thermal transport in one-dimensional (1D) classical many-body systems~\cite{Heat-1,Heat-2,Heat-3,Heat-4,Heat-5,Heat-6,Heat-7,Heat-8}: (i) Ballistic heat conduction in linear and nonlinear integrable systems~\cite{TypeA-1,TypeA-2,TypeA-3}; (ii) Superdiffusive heat conduction in nonlinear, nonintegrable systems with conserved momentum~\cite{TypeB-1,TypeB-2,TypeB-3}; and (iii) Normal (diffusive) heat conduction in nonlinear, nonintegrable systems without momentum conservation~\cite{NonInt-1,NonInt-2,NonInt-3,Typec-1,Typec-2,Typec-3}.

Although the above general scenario seems rather well established, recent studies on classical and quantum spin integrable chains have yielded a lot of unexpected findings~\cite{Spin-1,Spin-2,Spin-3,Spin-4,Spin-5}. Despite possessing a considerable number of conserved quantities, these systems can exhibit transport behaviors akin to those observed in fully nonintegrable systems. For instance, both diffusive and Kardar-Parisi-Zhang (KPZ)~\cite{KPZ} superdiffusive spin transport phenomena have been observed in classical and quantum integrable systems~\cite{Spin-4, Spin-5}. These surprising results, in turn, raise an intriguing question regarding classical many-body lattice systems: Is there a nonintegrable non-spin system that still deviates from the aforementioned general scenario?

In this Letter we present the first observation of ballistic thermal transport in a nonintegrable classical many-body system. By appropriately incorporating long-range interactions into a classical oscillator system, we systematically explore all the characteristic manifestations of ballistic transport within the system. This discovery thus challenges the prevailing wisdom that ballistic transport is exclusive to integrable systems, suggesting that the search for unexpected transport is not limited to spin systems. In principle, our work opens up exciting new directions for future research, including unraveling the underlying mechanisms responsible for ballistic transport in nonintegrable systems and exploring a comprehensive spectrum of transport behaviors across a broader range of systems.

\emph{Model}.---We consider a one-dimensional classical many-body Hamiltonian lattice of $N$ particles under the Born-von Karman periodic boundary conditions~\cite{BornvonKarman}, which is described by
\begin{equation}
H=\sum_{i}^N \left\{ \frac{p_{i}^{2}}{2}+ \frac{1}{2} (x_{i+1}-x_i)^2+ \sum_{r=1}^{\frac{N}{2}-1} \frac{\beta}{4 r^2} \left[x_{i+r}-(-1)^r x_i\right]^4 \right\}.
\label{HH}
\end{equation}
Here, $x_{i}$ and $p_{i}$ represent two canonically conjugated variables with $i$ denoting the particle index. The parameter $\beta > 0$ denotes the strength of nonlinearity, while $r$ controls the interaction range of long-range nonlinear interactions. All other relevant quantities, such as particle mass and lattice constant, are dimensionless and set to unity. Notably, this system exhibits a unique feature in its nonlinear interaction term due to the presence of $(-1)^r$, which may correspond to the nonlinear forces with alternating charges in certain 1D charged systems~\cite{Saito2017}.

We note here that the initial proposal of this system was to validate Peierls's hypothesis, which states that only umklapp processes can induce thermal resistance while normal processes cannot~\cite{Doi2022}. However, in our study, we employ this classical many-body system to address a distinct fundamental issue---the interplay between nonintegrability and ballistic thermal transport. In contrast to the weakly nonlinear case investigated in~\cite{Doi2022} (with $\beta=0.1$), we focus on a strongly nonlinear system with $\beta=1$, allowing us to emphasize the nonintegrable nature of the system. Consequently, we clearly demonstrate that even a nonintegrable system can exhibit ballistic thermal transport.
\begin{figure}[!t]
\vskip-0.4cm
\includegraphics[width=7cm]{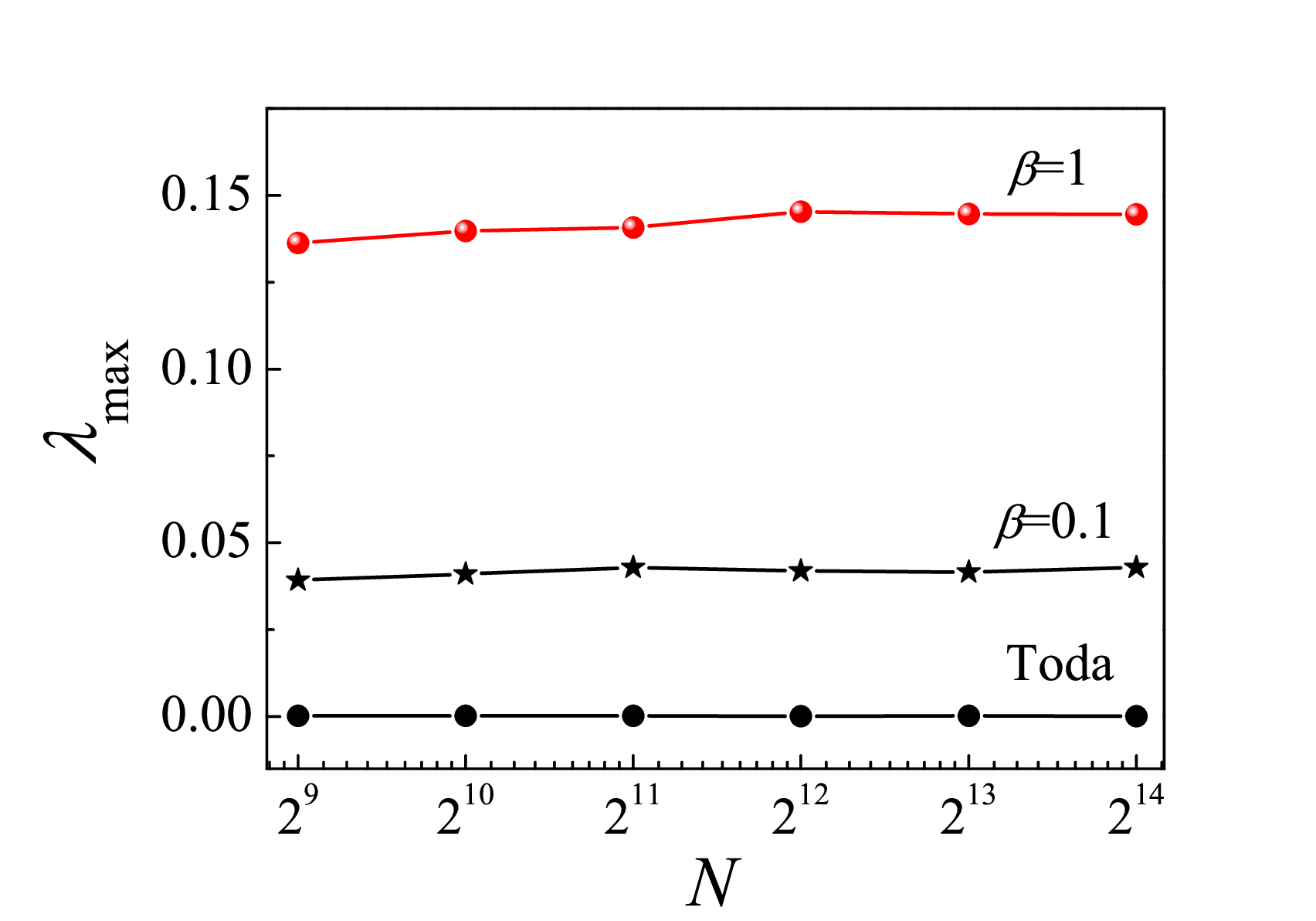}
\vskip-0.3cm
\caption{$\lambda_{\rm max}$ vs system size $N$ for a fully integrable nonlinear Toda system and our nonintegrable nonlinear systems with $\beta=0.1$ and $1$ (from bottom to top).} \label{fig:1}
\end{figure}

\emph{Nonintegrability}.---We first clarify that the system described by Hamiltonian~(\ref{HH}) with $\beta=1$ exhibits genuine nonintegrability, as evidenced by a non-vanishing maximal Lyapunov exponent $\lambda_{\rm max}$ ($>0$) (see Fig.~\ref{fig:1}), computed using the standard Benettin-Galgani-Strelcyn technique~\cite{Ly}. For comparison, we also plot the fully integrable Toda chain \{its Hamiltonian is $H=\sum_i^N p_i^2/2+ \exp[-(x_{i+1}-x_i)]+(x_{i+1}-x_i)-1$\} with $\lambda_{\rm max}=0$ and the system with $\beta=0.1$. It is evident that $\lambda_{\rm max}$ for $\beta=1$ surpasses those of both the Toda chain and the weakly nonlinear system, indicating a higher positive value. This trend remains consistent even when increasing the system size. Therefore, it unequivocally rules out any possibility of integrable dynamics in this strongly nonlinear regime.
\begin{figure}[!t]
\vskip-0.2cm
\includegraphics[width=9cm]{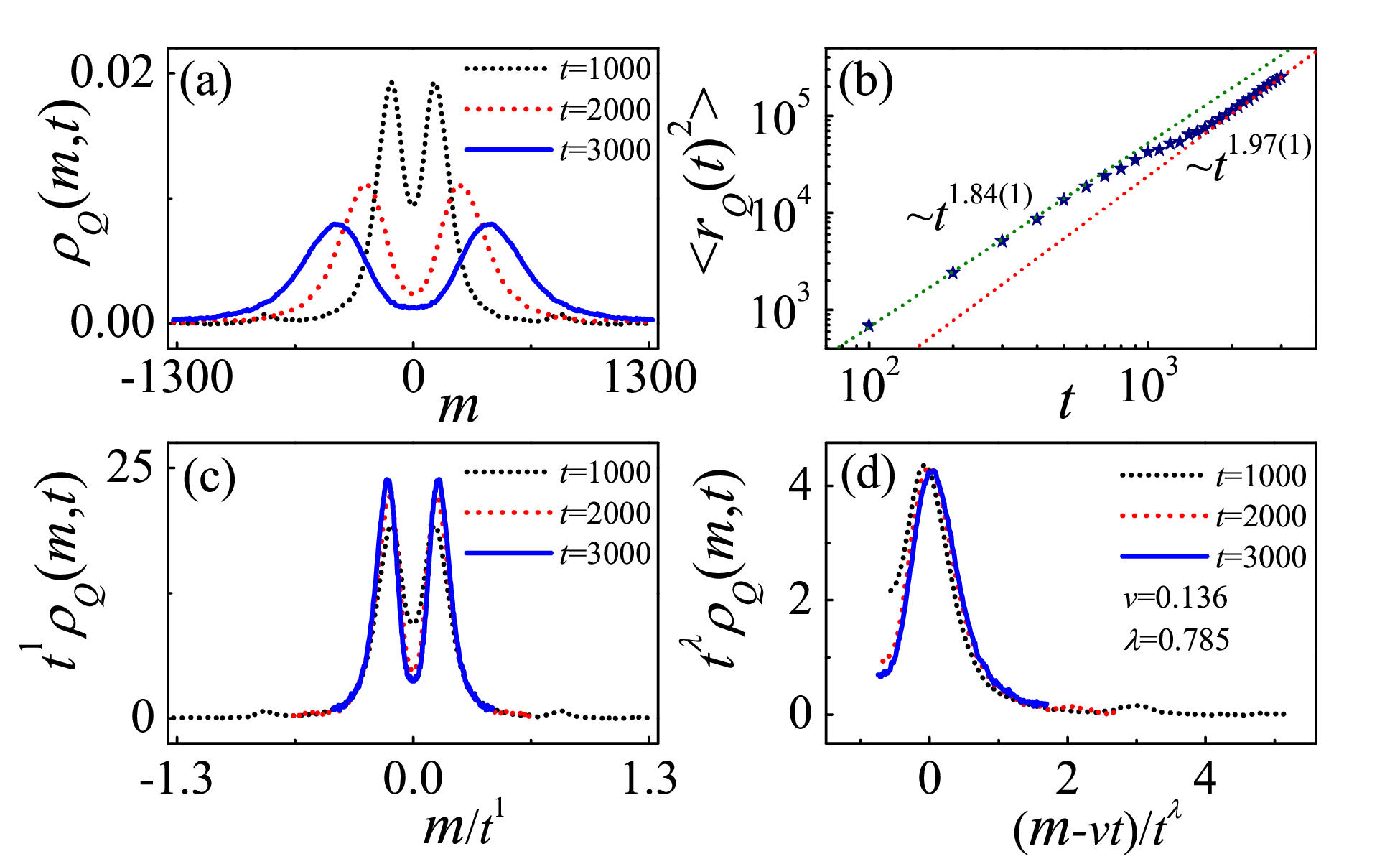}
\vskip-0.3cm
\caption{The ballistic scaling of equilibrium dynamical correlations: (a) the equilibrium spatiotemporal correlation function of thermal energy $\rho_{Q}(m,t)$ for several times $t$; (b) the mean squared displacement of $\rho_{Q}(m,t)$, i.e., $\langle r_Q(t)^2 \rangle$ vs $t$ showing ballistic transport in a long time limit; (c) $\rho_{Q}(m,t)$ for different times are scaled ballistically; (d) the ballistic front regions after the scaling transformation~(\ref{Scaling}).} \label{fig:2}
\end{figure}

\emph{Ballistic scaling of equilibrium dynamical correlations}.---Knowing the nonintegrability, we next study the heat diffusion property of the system. To capture this, we utilize the equilibrium (under temperature $T=1$ in a canonical system) spatiotemporal correlation function (see Refs.~\cite{Xiong2020,Xiong2022} for detailed information) of the local thermal energy $Q_l(t)$:
\begin{equation}
\rho_{Q}(m,t)=\frac{\langle \Delta Q_{l+m}(t) \Delta Q_{l}(0) \rangle}{\langle \Delta Q_{l}(0) \Delta Q_{l}(0) \rangle}.
\end{equation}
Here $Q_l(t)= E_l(t)-\frac{(\langle E \rangle + \langle F\rangle) g_l(t)}{\langle g \rangle}$, and due to translational invariance, the correlation only depends on the relative distance $m$. The notation $\langle \cdot \rangle$ denotes spatiotemporal averaging. The coarse-grained bin's number for computing $Q_l(t)$ is denoted by label $l$, and each bin contains $n=8$ particles. In the definition of $Q_l(t)$, we have particle number density as $g_l(t)$, energy density as $E_l(t)=\sum_{k \in l} E_k(t)$ (where each term within the sum corresponds to $\frac{p_{k}^{2}}{2}+ (x_{k+1}-x_k)^2/2+\frac{1}{4 r^2} [x_{k+r}-(-1)^r x_k]^4$, and it sums over all particles within bin $l$), and pressure density as $F_l(t)$. These densities can be evaluated by calculating the number of particles $g_l(t)$, total energy $E_l(t)$, and pressure $F_l(t)$ exerted on each bin at different time points. It should be noted that since our system possesses a symmetric potential, we have $\langle F\rangle = 0$.

To calculate $\rho_{Q}(m,t)$, we consider a total system size of $N=4096$, ensuring that an initial energy fluctuation located at the center can propagate up to time $t=3000$. The velocity-Verlet algorithm~\cite{Verlet} is employed with a small time step of $0.01$ for system evolution. In order to expedite our computations, we utilize a Fast Fourier Transform~\cite{FFT} algorithm. We emphasize that these computations are more challenging than traditional long-ranged systems due to the presence of the $(-1)^r$ term. Finally, an ensemble size of approximately $8 \times 10^9$ is utilized.

Figure~\ref{fig:2}(a) depicts $\rho_Q(m,t)$ for three long times showing ballistically moving peaks of the local thermal energy fluctuations. This provides an initial visual indication of ballistic thermal transport. To further verify this, Fig.~\ref{fig:2}(b) displays the mean squared displacement (MSD), $\langle r_Q(t)^2 \rangle$, calculated by
\begin{equation}
\langle r_Q(t)^2 \rangle=\sum_{m=-N/2}^{N/2} m^2 \rho_Q(m,t)
\end{equation}
versus time step $t$. It shows that $\langle r_Q(t)^2 \rangle$ scales as $t^{1.84}$ for short times and as $t^{1.97}$ at long times, which is already close to the ballistic scaling of $\langle r_Q(t)^2 \rangle \sim t^2$, taking into account numerical errors. Therefore, we argue that in our system, ballistic transport holds true in the long-time limit. The observed smaller scaling exponent of MSD at short times ($t\leq 1000$) is due to two small side peaks [see dotted line at $t=1000$ in Fig.~\ref{fig:2}(a)] and insufficient separation between two major moving peaks. Indeed, when performing the ballistic scaling transformation $t^{1} \rho_Q(m,t) \simeq \rho_Q(m/t^{1},t)$ on $\rho_Q(m,t)$ in Fig.~\ref{fig:2}(c), there is not a perfect scaling collapse. Because the two major peaks move with constant velocity, we also investigate their broadening over time by applying
\begin{equation}
t^{\lambda} \rho_Q(m,t) \simeq \rho_Q[(m-vt)/t^{\lambda},t]
\label{Scaling}
\end{equation}
with $v=0.136$ (subsonic speed) and scaling exponent $\lambda=0.785$. Interestingly, it exhibits non-Gaussian and non-KPZ behavior~\cite{KPZ}, indicating both uniqueness of this system and its distinctive ballistic behavior compared to fully integrable systems.
\begin{figure}[!t]
\vskip-0.2cm
\includegraphics[width=7cm]{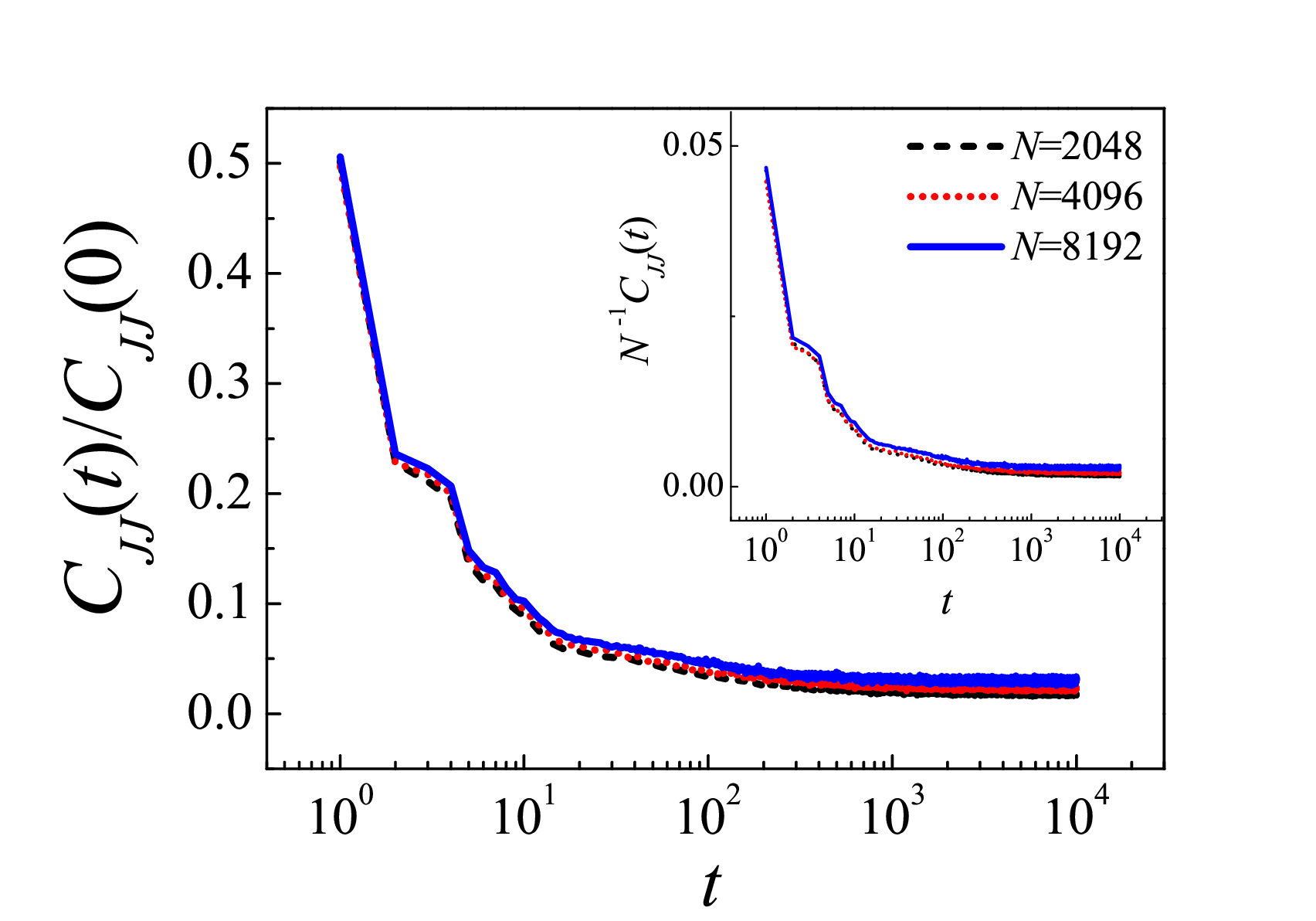}
\vskip-0.3cm
\caption{The equilibrium (the averaged temperature $T=1$) heat current autocorrelation $C_{JJ} (t)$ versus $t$ for several system sizes. Here for a better visualization, the values of $C_{JJ}(t)$ have been rescaled and divided by $C_{JJ} (0)$. The inset shows the same data but rescaled by $N^{-1}$.} \label{fig:3}
\end{figure}

\emph{Size-independent energy current}.---Another convincing evidence supporting ballistic transport can be obtained by investigating the system's equilibrium heat current autocorrelation $C_{JJ}(t)=\langle J_{\rm tot} (t) J_{\rm tot}(0) \rangle$. However, defining and computing the total heat current $J_{\rm tot}(t)$ in our system is not straightforward due to the presence of a specific term $(-1)^r$, which distinguishes it from conventional long-ranged systems~\cite{Xiong2020,Xiong2022}. We provide the definition in Supplementary Material (SM)~\cite{SM}.

Figure~\ref{fig:3} depicts $C_{JJ}(t)/C_{JJ}(0)$ for a long time step up to $t=10^4$ for several large system sizes. The spatiotemporal average of $C_{JJ}(t)$ is computed using a similar approach as $\rho_Q(m,t)$ under the same averaged temperature $T=1$. The collapse observed for different $N$ strongly suggests the size-independent energy current. Additionally, the inset reveals a scaling property $N^{-1}$, providing further evidence for ballistic transport in our system. Notably, we observe a finite nonzero plateau for $C_{JJ}(t)$ at $t \geq 10^3$, with values approximately equal to 3.35 ($N=2048$), 8.45 ($N=4096$), and 22.79 ($N=8192$) across all considered system sizes. This behavior differs from that exhibited by conventional fully integrable Toda systems where the plateau appears much earlier~\cite{NonInt-3}, but aligns with our observation that ballistic transport occurs over a relatively long timescale for this nonintegrable system. Such characteristics may represent unique aspects enabling ballistic transport in nonintegrable systems.
\begin{figure}[!t]
\vskip-0.2cm
\includegraphics[width=8.8cm]{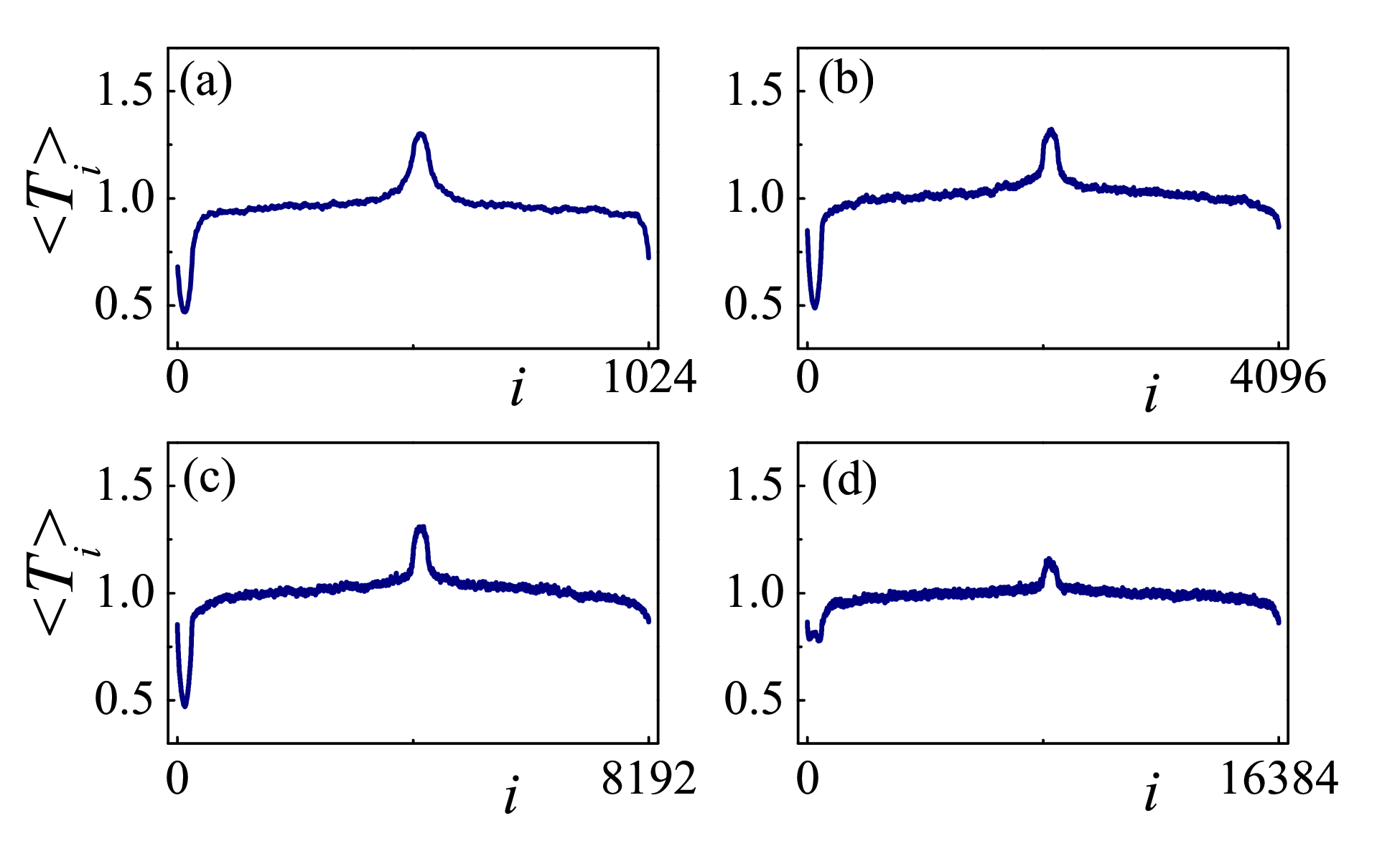}
\vskip-0.3cm
\caption{Temperature profiles for four typical large system sizes (the averaged temperature $T=1$).} \label{fig:4}
\end{figure}

\emph{Flat bulk temperature profile}.---Our third compelling evidence for ballistic thermal transport in the system is demonstrated by the absence of a temperature gradient. To achieve this, we utilized the ``reverse nonequilibrium molecular dynamics (RNEMD)'' method~\cite{RNEMD,Xiong2012} to generate the temperature profile due to dealing with a long-ranged interacting system. In this regard, the traditional approach~\cite{TypeB-1} for inducing a temperature profile encounters some challenges~\cite{Xiong2020}. The RNEMD approach was initially proposed for monoatomic fluids~\cite{RNEMD}, and in 2012, we presented an extensive implementation applied to 1D lattice systems~\cite{Xiong2012}. For its application to 1D long-ranged systems similar to here, we refer to~\cite{Xiong2020}.

In Fig.~\ref{fig:4} the temperature gradient in our system is vanishing for all considered system sizes. This flat bulk temperature profile thus further confirms the ballistic thermal transport in our system. Interestingly, Figs.~\ref{fig:4}(a)-(d) also reveal that the larger the system size, the flatter the bulk temperature profile, indicating that this unique ballistic thermal transport phenomenon occurs over longer timescales and in larger systems, which is consistent with the observations in Figs.~\ref{fig:2} and ~\ref{fig:3}.
\begin{figure}[!t]
\includegraphics[width=8.3cm]{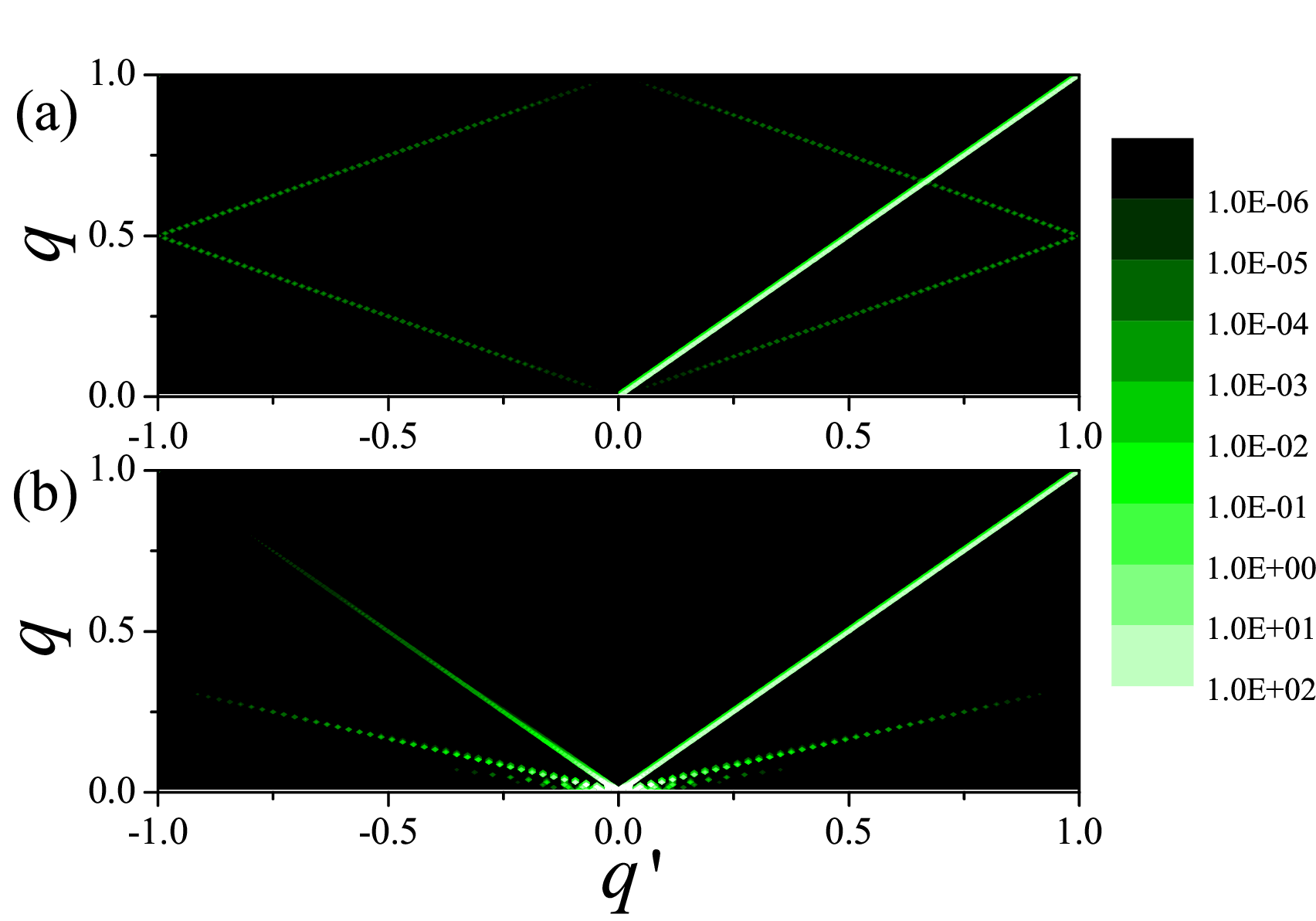}
\vskip-0.2cm
\caption{Normal mode excitations ($q'$, $x$-axis) from externally perturbed modes ($q$, $y$-axis) on a fully integrable Toda system (a) and our nonintegrable system with $\beta=1$. The scale on the right represents the ratio of the energy of the corresponding modes to that of the initial perturbed modes.} \label{fig:5}
\end{figure}

\emph{Possible underlying mechanisms}.---We finally explore the possible underlying mechanisms of the observed ballistic transport in our nonintegrable system. Noting that in a previous study~\cite{Doi2022}, the absence of umklapp processes in the weakly nonlinear regime of the system has been argued, here we will first investigate phonon scattering within the system and then elucidate which excitations contribute to the ballistic thermal transport in our nonintegrable system. Toward the first end, we employ the method proposed in~\cite{Phononscatter} to deduce phonon scattering from normal mode excitations. Figure~\ref{fig:5} illustrates the excited normal modes ($x$-axis) resulting from perturbing a single mode ($y$-axis) in both a fully integrable Toda system and our $\beta=1$ system. To present this, for both systems initially all the particles are at their equilibrium positions without any kinetic energy, then a single mode ($q$) is perturbed (excited) first, and the resultant excited normal modes ($q'$), which are measured by the magnitude of the modal energies, are tracked during a short observation time window of $0.1$ units. For details of the approach, we refer to~\cite{Phononscatter}. As depicted, for the Toda system [Fig.~\ref{fig:5}(a)], umklapp processes occur ($q+q=2q$) since the line of $q'=2q$ (due to the cubic anharmonicity) appears and this line can reach the boundary of Brillouin zone (BZ). However, for our system [Fig.~\ref{fig:5}(b)], even the line of $q'=3q$ (due to the quartic anharmonicity) corresponding to the four phonons processes ($q+q+q=3q$) emerges, only the perturbed mode itself ($q'=q$) can reach BZ's boundary---indicating that umklapp processes are indeed absent at least during this short-time period---and further confirming distinctions between our nonintegrable system and a fully integrable Toda one.
\begin{figure}[!t]
\includegraphics[width=8.3cm]{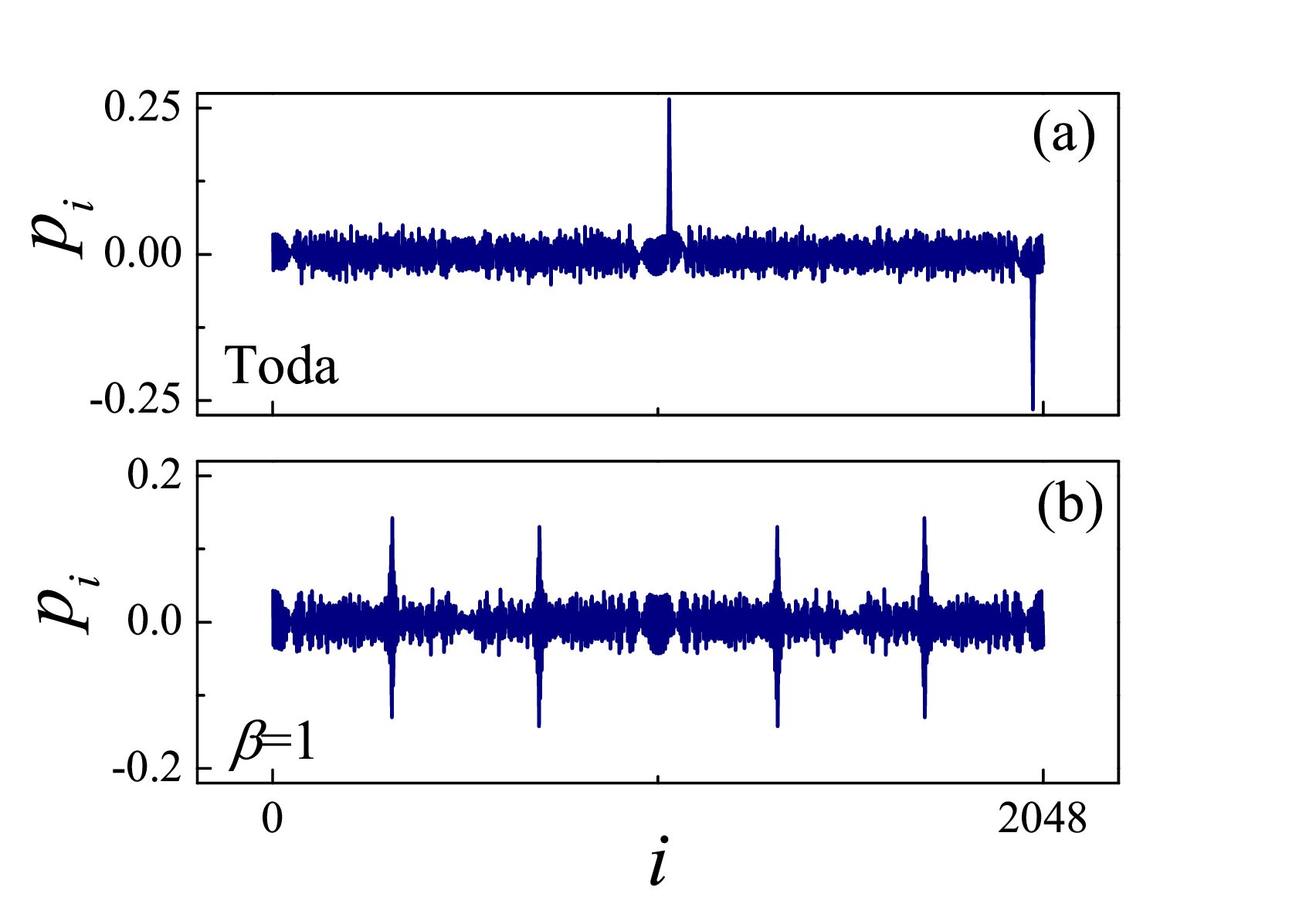}
\vskip-0.2cm
\caption{Snapshots of particle's momentum $p_i$ versus $i$ at a given long time $t=2000$, after initially applying two kicks at $i=1$ and $i=1025$ with momenta $p_1=-1$ and $p_{1025}=1$ (corresponding to a averaged temperature of $T=1$), for the fully integrable Toda system and our nonintegrable system with $\beta=1$.} \label{fig:6}
\end{figure}

We subsequently investigate the microscopic excitations of the system in its long-time behavior. To accomplish this, we conduct such a numerical experiment for $N=2048$: Initially, at time $t=0$, when all particles are in their equilibrium positions (the system is not thermalized as well), we apply two kicks to particles located at $i=1$ and $i=1025$ with momenta $p_1=-1$ and $p_{1025}=1$, respectively (corresponding to an average temperature of $T=1$). Subsequently, we meticulously analyze the evolution of the system's dynamics. A visually engaging animation illustrating this excitation dynamics for a time lag of $\delta t=10$, depicting each particle's momentum ($p_i$) as a function of $i$, can be found in SM~\cite{SM}. The relevant snapshots at a long time $t=2000$ are presented in Fig.~\ref{fig:6}. As evident from both SM and Fig.~\ref{fig:6}, there exists a clear distinction between integrable and nonintegrable systems: solitons account for ballistic transport in the Toda system [as observed by peaks in Fig.~\ref{fig:6}(a)], while our nonintegrable system with $\beta = 1$ exhibits moving discrete breathers (DBs). This finding suggests that even nonintegrable systems characterized by moving DBs can still display ballistic thermal transport and integrability is not necessary for ballistic transport.

\emph{Conclusion}.---It has long been believed that ballistic transport is a unique property of integrable systems, resulting from the presence of numerous conserved quantities. In contrast, nonintegrable systems, lacking sufficient conservation laws, are expected to exhibit either anomalous superdiffusive or normal diffusive transport. Surprisingly, our findings demonstrate that even in the absence of integrability, it is possible to achieve ballistic thermal transport by constructing a system with moving DBs as its primary microscopic excitations. These intriguing results are supported by all the typical indicators of ballistic transport in the long-time limit. This discovery challenges our current understanding of the relationship between integrability and transport behavior and significantly expands our comprehension of thermal transport in low-dimensional materials. It also suggests new possibilities for engineering materials with ideal ballistic thermal properties driven by traveling DBs and enhances our understanding of fundamental physical processes in complex systems.

\begin{acknowledgments}
J.W. is supported by NNSF (Grant No. 12105122) of China; D.X. is supported by NNSF (Grant No. 12275116) of China, NSF (Grant No.
2021J02051) of Fujian Province of China, and the start-up fund of Minjiang University.
\end{acknowledgments}


\end{document}